\documentclass[11pt]{article}
\usepackage{amsmath}
\usepackage{amssymb}
\usepackage{hhline}
\usepackage[scale={.75,.75}]{geometry}
\usepackage{url}
\usepackage{caption}
\usepackage[numbers, sort&compress]{natbib}
\usepackage{epsfig}
\usepackage{multirow}
\usepackage{color}
\usepackage{verbatim}
\usepackage{nicefrac}
\usepackage{upgreek}
\usepackage{bbm}
\usepackage{setspace}

\usepackage{hyperref}% add hypertext capabilities

\newcommand{\Dm}{\Delta^-}
\renewcommand{\Re}{\text{Re}}
\renewcommand{\Im}{\text{Im}}
\newcommand{\Dp}{\Delta^+}
\newcommand{\Pm}{\Pi^-}
\newcommand{\Pp}{\Pi^+}

\newcommand{\wqi}{\omega_{\textbf{q}}(t_1)}
\newcommand{\q}{\textbf{q}}
\newcommand{\W}{\Omega}
\newcommand{\G}{\Gamma}
\newcommand{\tp}{t}
\newcommand{\fex}{\textit{e.g.}~}
\newcommand{\ie}{\textit{i.e.}~}
\newcommand{\TI}{\tau_\text{int}}

\begin{document}
\date{}

\title{
\begin{flushright}
{\small MPP-2012-3, TTK-12-03,TUM-HEP-857/12} %\hfill\mbox{}\\}
\end{flushright}
\vspace{0.5cm}
\Large{\bf The Boltzmann Equation from Quantum Field Theory}}
%\title{The Boltzmann Equation from Quantum Field Theory}  

\author{Marco Drewes$^{a,b}$, 
Sebasti\'an Mendizabal$^{c}$
and Christoph~Weniger$^{d}$ \\[3mm]
{\normalsize \it$^a$ Institut f\"ur Theoretische Teilchenphysik und
Kosmologie,}\\
{\normalsize \it RWTH Aachen, 52056 Aachen, Germany}\\
{\normalsize \it$^b$ Physik Department T70, Technische Universit\"at M\"unchen,} \\{\normalsize \it James Franck Stra\ss e 1, D-85748 Garching, Germany}\\
{\normalsize \it$^c$ Institut f\"ur Theoretische Physik, Goethe-Universit\"at,}\\
{\normalsize \it 60438 Frankfurt am Main, Germany}\\
{\normalsize \it$^d$ Max-Planck-Institut f\"ur Physik, F\"ohringer Ring 6,
80805 M\"unchen, Germany}
}
\maketitle
\begin{abstract}
  \noindent
We show from first principles the emergence of classical Boltzmann equations from relativistic nonequilibrium quantum field theory as described by the Kadanoff-Baym equations. Our method applies to a generic quantum field, coupled to a collection of background fields and sources, in a homogeneous and isotropic spacetime.
The analysis is based on analytical solutions to the full Kadanoff-Baym equations, using the WKB approximation. 
This is in contrast to previous derivations of kinetic equations that rely on similar physical assumptions, but obtain approximate equations of motion from a gradient expansion in momentum space.
We show that the system follows a generalized Boltzmann equation whenever the WKB approximation holds. 
The generalized Boltzmann equation, which includes off-shell transport, is valid far from equilibrium and in a time dependent background, such as the expanding universe.
\end{abstract}

%\pacs{pacs}

\section{Introduction}%
Nonequilibrium phenomena play a crucial role in many areas of physics,
including the early history of the universe, heavy ion collisions, condensed
matter physics and quantum information.  In the era of precision cosmology and
with the arrival of the LHC and RHIC experiments, in particular the first two
applications, which require a relativistic description, have gained
considerable interest. Transport in nonequilibrium situations can often in
very good approximation be described by Boltzmann equations (BEs). These
assume that the system can be characterized by a number of distribution
functions for classical particles, which propagate freely between isolated
interactions and carry no memory of their history. However, the definition of
asymptotic states, on which the single particle description is based, is
ambiguous in a dense plasma. What is more, the standard BEs by construction
cannot describe memory and off-shell effects or quantum coherence. Usually
these effects are treated by effective kinetic equations of the Boltzmann
type~\cite{CalzettaHu,kb62,Bogo, Danielewicz:1982kk, Chou:1984es, Sigl:1992fn,Canetti:2012vf,Jeon:1995zm,
Greiner:1998vd, Calzetta:1999xh, Boyanovsky:1999cy, Buchmuller:2000nd,Cline:2000nw,
Arnold:2000dr, De Simone:2007rw, Hohenegger:2008zk, Anisimov:2008dz,Drewes:2010pf,Garny:2009ni,Garny:2009rv,
FillionGourdeau:2006hi, Cirigliano:2009yt, Beneke:2010dz,Anisimov:2010dk,Herranen:2010mh,Prokopec:2003pj,hep-ph/0512155, Calzetta:1986cq,Garbrecht:2011xw,Gautier:2012vh,Boyanovsky:2004dj,Ivanov:1999tj,Blaizot:2001nr}, \ie by a set
of first order differential equations for generalized distribution functions
that are local in time. As the above issues are conceptual, their range of
validity and possible corrections cannot be determined within a framework of
BEs and require a derivation from first principles. 

The full equations of motion of nonequilibrium quantum field theory, on which
first-principle derivations of the BEs are usually based, are known as
Kadanoff-Baym equations (KBEs)~\cite{kb62} \footnote{The KBE are equations of motion for correlation functions. Alternatively one can use equations of motion for the fields themselves as starting point, c.f. \cite{Gautier:2012vh} and references therein for a detailed comparison.}. These equations, being coupled
second order integro-differential equations, are considerably more complicated
than BEs. 
Most approaches to establish a connection between
out-of-equilibrium quantum fields and kinetic equations make a number
of approximations on the KBEs \textit{before} they are solved (\fex
Refs.~\cite{Buchmuller:2000nd,kb62,Calzetta:1999xh,Chou:1984es,Calzetta:1986cq,Prokopec:2003pj, hep-ph/0512155, De
Simone:2007rw, Cirigliano:2009yt,Cline:2000nw, Beneke:2010dz, Hohenegger:2008zk,Garny:2009rv,Herranen:2010mh,Ivanov:1999tj,Blaizot:2001nr}). 
Starting point is usually a gradient expansion, performed in Wigner-space,
which provides a consistent approximation scheme when a separation of scales is realized in the system. 
Common additional simplifications include a close-to-equilibrium assumption for all fields, the quasiparticle
approximation and the Kadanoff-Baym ansatz for correlation functions. 
However, the Wigner space method as such does not rely on these additional assumptions if the gradient expansion is performed consistently, which may require resummations \cite{Garbrecht:2011xw}.

In this letter, we show how the
\textit{full} KBEs can be solved by using the Wentzel-Kramers-Brillouin (WKB)~\cite{WKB} method (for
earlier uses of the WKB method in a similar context see \fex
Refs.~\cite{Cline:2000nw, Prokopec:2003pj}). 
This approach avoids the Fourier transformation to Wigner space in relative time and uses what is sometimes called the {\it two times formalism}.
It is valid far from equilibrium and does not rely on an on-shell approximation or any other a priori assumption about the form of the correlation functions, such as the Kadanoff-Baym ansatz.
We illustrate our method for
a real scalar field, coupled to other fields, in a spatially homogeneous and isotropic background.
This choice is for transparency only; the derivation does not rely on assumptions 
about the spin and interactions of the field or background.
Though technically more difficult, the generalization to fermions with gauge interactions 
is straightforward.

\section{Nonequilibrium quantum field theory}%
We consider the dynamics of a real scalar field $\phi$ that is described by
relativistic quantum field theory. The field $\phi$ weakly couples to a
background, possibly containing many degrees of freedom whose dynamics is in
principle known and that we refer to as $\chi_i$. The Lagrangian reads
\begin{equation}\label{L}
  \mathcal{L}=\frac{1}{2}\partial^\mu\phi\partial_\mu\phi -
  \frac{1}{2}m(t)^2\phi^2-\phi\ \mathcal{O}[\chi_i,t]
  +\mathcal{L}_{\chi_i}\;,
\end{equation}
where $\mathcal{ O}[\chi_i,t]$ denotes a sum of generic combinations of fields $\chi_i$
with coefficients that may depend on time explicitly\footnote{
Our approach does not rely on the way $\phi$ couples to the bath as long as conditions 1)-3) specified in section \ref{sec:towBE} are fulfilled. In (\ref{L}) we chose a coupling that is linear in $\phi$ to obtain the simple explicit expressions (\ref{SelfEnergiesDef}) for the self-energies and to justify the time translation invariance of $\Pi^\pm$ in equations (\ref{solution})-(\ref{tsunami}) All other formulae and considerations remain valid for an arbitrary coupling between $\phi$ and other fields.
}.
$\mathcal{L}_{\chi_i}$ determines the dynamics of $\chi_i$ (we use
$\hbar=c=1$). We allow a time-dependent mass $m(t)$ to account for Hubble
expansion when interpreting $t$ as conformal time, 
the time-dependence of other operators is contained in $\mathcal{O}[\chi_i,t]$ and $\mathcal{L}_{\chi_i}$. 

In quantum physics, any thermodynamic system can be characterized by a density
operator $\varrho$. Knowledge of the density operator
allows to compute expectation values for all observables at all times.
The same information is contained in the set of all
$n$-point functions $\langle \phi(x_1) \cdots \phi(x_n)\rangle$ etc. of the fields, (with $\langle \cdots \rangle\equiv\text{tr}[\varrho \cdots]$). However, most quantities of practical interest for which one formulates a Boltzmann equation
can be expressed in terms of one- and two-point functions; this includes the
energy-momentum tensor and charge densities. It is, therefore, usually
sufficient to track the time evolution of these. 

An out-of-equilibrium quantum field has two independent connected two-point
functions. In case of $\phi$ they are conveniently chosen as
\begin{eqnarray}
  \label{CorrDef}
  \Delta^{-}(x_1,x_2)\equiv i\langle[\phi(x_1),\phi(x_2)]\rangle\;,\\
  \nonumber
  \Delta^{+}(x_1,x_2)\equiv
  \frac{1}{2}\langle\{\phi(x_1),\phi(x_2)\}\rangle\;,
\end{eqnarray}
with the obvious symmetry relations
$\Delta^\pm(x_2,x_1)=\pm\Delta^\pm(x_1,x_2)$. Here $[,\!]$ and $\{,\!\}$ are
commutator and anti-commutator, respectively. $\Delta^{-}(x_1,x_2)$ is known
as \textit{spectral function} and basically encodes information about the
spectrum of resonances in the thermodynamic description, 
which may differ from the spectrum in vacuum.  The \textit{statistical propagator}
$\Delta^{+}(x_1,x_2)$ carries information about the occupation numbers of
different modes.
We will in the following derive the quantum field theory analogue to the classical particle distribution function from the statistical propagator.  We have in mind applications in cosmology and
restrict the analysis to spatially homogeneous and isotropic systems. Then, the correlation
functions only depend on relative spatial coordinates
$\textbf{x}_1-\textbf{x}_2$ etc., and it is convenient to perform a spatial
Fourier transform in these coordinates, yielding functions like
$\Delta^\pm_{\textbf{q}}(t_1,t_2)\equiv\int d^3(\textbf{x}_1-\textbf{x}_2)
e^{-i\textbf{q}(\textbf{x}_1-\textbf{x}_2)}\Delta^\pm(x_1,x_2)$.

In a general out-of-equilibrium system the two-point functions
$\Delta^\pm_\q(t_1,t_2)$ have to be found as solutions to the KBEs
\begin{align}
  \left(\partial_{t_1}^2\!\! +\! \wqi^2 \right) \Dm_{\textbf{q}}(t_1,t_2) &=
  -\!\int_{t_2}^{t_1}\!\!\! dt'  \ \Pm_{\textbf{q}}(t_1, t')
  \Dm_{\textbf{q}}(t', t_2)\label{kb1}\\
  \left(\partial_{t_1}^2\!\! +\! \wqi^2 \right) \Dp_{\textbf{q}}(t_1,t_2)
  &= -\!\int_{t_i}^{t_1}\!\!\! dt'\ \Pm_{\textbf{q}}(t_1, t')
  \Dp_{\textbf{q}}(t', t_2)
  \nonumber\\ &\!\!\!\!\!\!\!\!\!\!\!\!\! 
  +\! \int_{t_i}^{t_2}\!\!\! dt'\ \Pp_{\textbf{q}}(t_1, t')
  \Dm_{\textbf{q}}(t', t_2)\label{kb2}\;,
\end{align}
where $\omega_\q(t)^2 \equiv m(t)^2 + \q^2$ (note that in equilibrium
$\Delta^\pm_\q$ would only depend on $t_1-t_2$), and $t_i$ denotes the initial
time of the system.  The KBEs can be derived within the Schwinger-Keldysh
formalism, see \fex \cite{Schwinger:1960qe, Keldysh:1964ud, Luttinger:1960ua,
Berges:2004yj, Calzetta:1986cq}. 
The first term on the RHS of (\ref{kb2}) is associated with non-Markovian (memory) effects while the second is often referred to as noise term.
The boundary conditions for $\Delta^-_\q$ are fixed by microcausality and canonical quantization for a real scalar field,
$\Delta^-_\q|_{t_1=t_2}=0$, $\partial_{t_1}\Delta^-_\q|_{t_1=t_2} =
-\partial_{t_2}\Delta^-_\q|_{t_1=t_2} =1$ and
$\partial_{t_1}\partial_{t_2}\Delta^-_\q|_{t_1=t_2}=0$. The boundary
conditions for $\Delta^+_\q|_{t_1=t_2=t_i}$ are determined by the physical
initial conditions of the system at time $t_i$. 
For simplicity we assumed Gaussian initial correlations for $\phi$, more general initial conditions are e.g. discussed in \cite{Garny:2009ni}.
Below, we will drop momentum indices $\q$ when possible.

The quantities $\Pi^\pm$ appearing in \eqref{kb1} and \eqref{kb2} are the
self-energies of $\phi$; in analogy to \eqref{CorrDef} they are at leading
order in $\mathcal{O}[\chi_i]$ given by
\begin{eqnarray}\label{SelfEnergiesDef}
  \Pi^{-}(x_1,x_2)&=&
  \langle\big[\mathcal{O}[\chi_i(x_1),t_t],\mathcal{O}[\chi_i(x_2),t_2]
  \big]\rangle\;, \\\nonumber
  \Pi^{+}(x_1,x_2)&=&-\frac{i}{2}\langle\big\{
  \mathcal{O}[\chi_i(x_1),t_1], \mathcal{O}[\chi_i(x_2),t_2]
  \big\}\rangle\;,
\end{eqnarray}
and contain information about the interaction between $\phi$ and the
background fields $\chi_i$. They can be calculated in terms of the two-point
functions of $\chi_i$ within the 2PI formalism (see \fex
Ref.~\cite{Luttinger:1960ua, Berges:2004yj} for details).

\section{Deriving the Boltzmann Equation}\label{sec:towBE}%
We will discuss the emergence of a description of $\phi$ in terms of effective
kinetic equations by using analytical solutions of the full KBEs that are
found with the WKB method. To this end, we make the following assumptions (we
also send $t_i\to-\infty$, effects of finite $t_i$ are discussed below):

1) The self-energies $\Pi^\pm(t_1,t_2)$ are damped with respect to the
relative time $|t_1-t_2|$, approaching zero for $|t_1-t_2|\gtrsim \TI$, where
we introduced the \textit{interaction time} $\TI$. Here, $\TI$ can be
considered as definition for the duration of \fex scattering events. Then evaluating one-sided Fourier transforms of the self-energies with respect to relative time,
\begin{equation} 
  \label{eqn:Dms}
  \widetilde\Pi^\pm(t, \omega)\equiv\int_0^\infty dz\ e^{i\omega z}\
  \Pi^{\pm}(t,t-z)\;,
\end{equation}
practically does not require knowledge of the system in the distant past $z\gg \TI$.
In equilibrium the minus-component would correspond to the common retarded
self-energy, $\widetilde\Pi^-(t,\omega) = \Pi^R(\omega)$.

2) We assume that for fixed time $t$ the pole structure of $(\omega^2 -
\omega_\q^2(t)- \widetilde\Pi^-(t, \omega ))^{-1}$ is dominated by the root
$\omega=\widehat \W_t\equiv\W_t - \frac i2 \Gamma_t$, with
\begin{equation} 
  \W_t\equiv \sqrt{\omega_\q^2(t) + \Re \widetilde\Pi^-(t, \widehat \W_t)}
  \;,\;\;
  \G_{t} \equiv - \frac{\Im\widetilde\Pi^-(t,
  \widehat\W_t)}{\W_t }\label{Gammadef2}\;.
\end{equation} 
Here $\W_t$ and $\G_t$ are the energy and damping rate of the
$\phi$-resonance, and we assume \emph{weak damping} with $\G_t\ll \W_t$,
a generic consequence of weak coupling.  
The generalization to the case with more roots (interpreted as collective
$\phi$ excitations) is straightforward. 

3) Three time scales are relevant for our discussion: The above
interaction time $\TI$, the damping rate $\Gamma_t$ of the field $\phi$ \footnote{The physical damping rate is obtained after renormalization. This usually amounts to multiplication of $\Gamma_t$ by a wave function renormalization constant \cite{Boyanovsky:2004dj,Anisimov:2008dz}.}, and
the characteristic rate $H\sim \dot\G_t/\G_t\sim\dot\W_t/\W_t$ with which the
field $\phi$ changes its properties (\fex due to a temperature change of the
fields $\chi_i$ in an expanding universe; then $H$ denotes the
Hubble rate). The main assumption, underlying all derivations of kinetic
equations, is the separation of time scales, 
which defines the small parameters that control the WKB approximation:
\begin{equation}\label{smallnesses}
  \G_t\ \ll \TI^{-1} \quad \text{and}\quad \dot\G_t/\G_t,\ \dot\W_t/\W_t\ \ll\
  \TI^{-1}\;.
\end{equation}
The first condition implies that the duration of one individual collision is
shorter than 
the average time between different collisions; the second
condition implies that the field does not significantly change its properties
during a single collision. The often-used gradient expansion of the KBE 
relies on the smallness of the same parameters (\ref{smallnesses}).
 
The damping of the self-energies $\Pi^\pm(t_1,t_2)$ required in assumption 1) is often governed by a power law.
In contrast to an exponential, there is no uniquely defined scale associated with a power law. Hence, the definition of the time $\TI$ beyond which the suppression is ``sufficient'' is not unique and depends on the accuracy one demands.
Throughout, we will not specify the interaction time $\TI$, which
has to be found on a case by case basis, and leave it as a free parameter 
\footnote{A more strict notion of locality can be imposed on the integrals on the RHS of (\ref{kb1}) and (\ref{kb2}) \cite{Gautier:2012vh}. Though weaker than the locality condition on $\Pi^\pm$, it is usually sufficient for what follows.}.
Exceptionally small temporal damping is in general associated with sharp features in the Fourier
transform $\widetilde\Pi^\pm(t, \omega)$ and corresponds to threshold effects or resonant phenomena.

The localization in time occurs for different physical
reasons: If all fields that appear inside $\Pi^\pm$ reside in a thermal bath and are weakly coupled,
their propagators are exponentially damped with the thermal damping rate; as a consequence $\Pi^\pm$ are also localized.  More generally,
when considering the scattering of particles, the duration of a scattering
event typically is related to the de Broglie wavelength $\sim 1/M$ of the interacting
particles with energies $\sim M$; formally the localization arises
from the momentum integrals inside of loops. Finally, virtual particles can only exist for times $\Delta t \lesssim M^{-1}$ as
allowed by the uncertainty principle. 

Under the above assumptions, the first KBE \eqref{kb1} can be approximately
solved by using the WKB method~\cite{WKB}, which is formally an expansion of
the solution to \eqref{kb1} in terms of the Planck constant
$\hbar$~\cite{Prokopec:2003pj}. At second order in $\hbar$, we find 
\begin{align}
  \Dm(t_1,t_2) = \frac{
  \sin\left(\int_{t_2}^{t_1}dt'\ \W_{t'}\right)
  e^{-\frac12\left|\int_{t_2}^{t_1}dt'\ \G_{t'}\right|}
  }{\sqrt{\W_{t_1}\W_{t_2}}}
  \label{eqn:Ansatz}.
\end{align}
For small time separations, $|t_1-t_2|\lesssim \TI$, the time-dependence of
$\widetilde \Pi^-(t,\omega)$ is negligible, and \eqref{eqn:Ansatz} can be
obtained by a Laplace transformation as in~\cite{Anisimov:2008dz,Anisimov:2010dk}. For
$|t_1-t_2|\gtrsim\TI$, we illustrate the derivation of the WKB result by
splitting $\Dm$ into positive and negative frequency modes, $\Dm
\equiv\frac{i}{2} (\Dm_{(+)} - \Dm_{(-)})$.  Inserting the functions
$\Dm_{(\alpha)}(t_1,t_2)$ with $\alpha=\pm$ into \eqref{kb1} yields
(for $t_1>t_2$)
\begin{align}
  \label{eqn:step1}
  \omega_\q^2(t_1) &-\W^2_{t_1} -i\alpha\W_{t_1}\G_{t_1} 
  \\ & \nonumber
  \simeq-\int^{\infty}_0 dz\ \Pi^-(t_1, t_1-z) 
  e^{\int_{t_1}^{t_1-z}dt' (i\alpha \W_{t'}-\frac12\G_{t'})}\;,
\end{align}
where we neglected terms suppressed by \eqref{smallnesses} or $\G_t/\W_t$;
terms containing $\dot\Omega_{t_1}$ cancel out. Since $\Pi^-(t_1,t')$
effectively vanishes for $|t_1-t'|\gtrsim\TI$ we extended the $z$-integration
limit formally to $\infty$. $\W_t$ and $\G_t$ are practically constant over
the support of $\Pi^-$, hence the $z$-integral equals $\Re\widetilde\Pi⁻(t_1,
\widehat \Omega_{t_1})-i \alpha\Im\widetilde\Pi⁻(t_1, \widehat\Omega_{t_1})$.
Then, the real and imaginary parts of \eqref{eqn:step1} reproduce the
definitions  \eqref{Gammadef2} of $\W_t$ and $\G_t$. The case $t_1<t_2$ is
treated analogously.

Knowledge of the spectral function $\Dm$ and the self-energies $\Pi^\pm$ are
enough to find a solution to the second KBE \eqref{kb2}; it is given by the
\textit{memory integral}
\begin{align}
  \Dp(t_1,t_2) \!= \!\!
  \int^{t_1}_{t_i}\!\!\! dt_1'\!\! \int^{t_2}_{t_i}\!\!\! dt'_2\ \Dm(t_1,t'_1)
  \Pp(t'_1,t'_2) \Dm(t'_2,t_2)\;,
  \label{MemInt}
\end{align}
where we allowed for a finite initial time $t_i$. This can be obtained by
using the initial conditions of $\Dm$ and the fact that $\Dm$ solves the first
KBE~\eqref{kb1}.

For convenience, we now split \eqref{MemInt} into three terms,
\begin{align}
  \Delta^+(t_1,t_2) = \mathcal{B}(t_1,t_2) +
  \mathcal{C}(t_1,t_2)+\mathcal{D}(t_1,t_2)\;.
  \label{eqn:split}
\end{align}
Here, $\mathcal{B}$ will correspond to the classical Boltzmann behavior,
whereas $\mathcal{C}$ and $\mathcal{D}$ will give corrections that can be
neglected within our approximations. The terms $\mathcal{B}$ and $\mathcal{C}$
are defined to contain all contributions to $\Delta^+$ that come from the time
integration over $t_1',t_2'<t_B\equiv \min(t_1,t_2)$ in \eqref{MemInt}, where we
defined the \emph{Boltzmann time} $t_B$. The term $\mathcal{D}$ contains the
remaining contributions that come from outside this region. For a self-energy
that is exactly local in time, $\Pi^+(t_1, t_2) \sim \delta(t_1, t_2)$, only
$\mathcal{B}$ and $\mathcal{C}$ would contribute, $\mathcal{D}$ would be
identically zero. In more general cases, $\mathcal{D}$ is is suppressed
by $\G_t \TI$ 
and remains negligible.

Now, $\mathcal{B}+\mathcal{C}$ is split such that $\mathcal{B}$ contains all
terms with equal-sign frequencies,
$\mathcal{B}\sim\Delta^-_{(\alpha)}\Delta^-_{(\alpha)}$; $\mathcal{C}$
contains the remaining opposite-sign terms,
$\mathcal{C}\sim\Delta^-_{(\alpha)}\Delta^-_{(-\alpha)}$.  Then, $\mathcal{C}$
can be written as
\begin{align}
  \nonumber
  \mathcal{C}(t_1,t_2)\!
  &=\!
  \frac{e^{i\int_{t_2}^{t_1}\!dt' \W_{t'}}
  e^{-\frac12\left|\int_{t_2}^{t_1}\!dt'\G_{t'}\right|}}
  {2\sqrt{\W_{t_1} \W_{t_2}}}\!\!\!
  \int^{t_B}_{-\infty}\!\!\!\!\!\! d\tau\ \! e^{-\!\int_{\tau}^{t_B}\!dt'(\G_{t'}+ 2i
  \W_{t'})} 
  \\ &\phantom{}\!\!\!\!\! \!\!\!\!\!\!\!\!\!\times
  \int^{\infty}_{0}\!\! dz
  \frac{\Pp(\tau, \tau-z)}{\sqrt{\W_{\tau}\W_{\tau-z}}} e^{\int_{\tau-z}^{\tau}dt'\!
  (i\W_{t'}-\frac12 \G_{t'})} +\text{h.c.}
\end{align}
The first term in the second line corresponds to $\widetilde\Pi^+(\tau,
\widehat \W_{\tau}^\ast)/\W_{\tau}$. Unless this expression is oscillating
with frequencies $\pm2\Omega_{\tau}$, the integral over $\tau$ is of the order
$\Im \widetilde\Pi^+(\tau,\widehat\W_{\tau}^\ast)/\Omega_{\tau}^2$ and hence
negligible. Finally, $\mathcal{B}$ can be written as
\begin{align}
  \nonumber
  \mathcal{B}(t_1,t_2)\!
  &=\!
  \frac{\cos\left(\int_{t_2}^{t_1}\!dt' \W_{t'}\right)
  e^{-\frac12\left|\int_{t_2}^{t_1}dt'\G_{t'}\right|}}
  {-2\sqrt{\W_{t_1} \W_{t_2}}}\!\!
  \int^{t_B}_{-\infty}\!\!\! d\tau e^{-\int_{\tau}^{t_B}dt' \G_{t'}} 
  \\ &\phantom{}\!\!\!\!\!\!\!\!\!\!\! \!\!\!\!\!\!\!\!\!\times
  \int^{\infty}_{0} \!\!dz\ \!
  \frac{\Pp(\tau, \tau-z)}{\sqrt{\W_{\tau}\W_{\tau-z}}} e^{\int_{\tau-z}^{\tau}dt'
  (i\W_{t'}-\frac12 \G_{t'})} + \text{h.c.}\;,
  \label{eqn:Bexp}
\end{align}
where again we obtain $\widetilde\Pi^+(\tau, \widehat
\W_{\tau}^\ast)/\W_{\tau}$ in the second line.  

From \eqref{eqn:Bexp}, and adopting the common definitions
\begin{align}
  2{\rm Re}\widetilde{\Pi}^+(\widehat \W_t^\ast )\equiv-\W_t \big(\G^>_{t}+\G^<_{t}\big)
  \; , \;\;
  \G_{t}\equiv\G^>_{t}-\G^<_{t}\;,
\end{align}
we obtain for the statistical propagator $\Dp$, up to terms that are
suppressed by $\Gamma_t/\Omega_t$ and $\G_t\TI$,
\begin{align}
  \label{eqn:DpStep}
  \Dp(t_1,t_2)
  &\simeq
  \frac{\cos\left(\int_{t_2}^{t_1}\!dt' \W_{t'}\right)
  e^{-\frac12\left|\int_{t_2}^{t_1}dt'\G_{t'}\right|}}
  {2\sqrt{\W_{t_1} \W_{t_2}}}
  \\ &\phantom{} \!\!\!\!\!\!\!\!\!\times \nonumber
  \underbrace{\int^{t_B}_{-\infty}\!\! d\tau
  \left(\G^>_{\tau}+\G^<_{\tau}\right)
  e^{-\int_{\tau}^{t_B}dt'\ (\G^>_{t'}-\G^<_{t'})}}_{\equiv1+2f(t_B)} \;.
\end{align}
We can now \textit{define} the suggestive quantity $f(t_B)$ as function of
$t_B$ as indicated in \eqref{eqn:DpStep}. Most importantly, from its very
definition it follows that $f(t_B)$ solves the kinetic equation
\begin{equation}\label{EffBE}
  \partial_{t_B} f(t_B) = \left( 1+f(t_B) \right)\G^<_{t_B} - f(t_B)\G^>_{t_B}\;,
\end{equation}
which is of first order and local in time, \ie has the properties of a generalized BE.
Then, our result for the statistical propagator becomes,
up to terms suppressed by \eqref{smallnesses} or $\G_t/\W_t$,
\begin{align}
  \Dp(t_1,t_2) \!=\! \frac{
  \cos\left(\int_{t_2}^{t_1}\!dt' \W_{t'}\right)\!
  e^{-\frac12\left|\int_{t_2}^{t_1}dt' \G_{t'}\right|}
  }{2\sqrt{\W_{t_1}\W_{t_2}}} \left(1\!+\!2f(t_B) \right)\;,
  \label{eqn:Dp2}
\end{align}
which makes explicit that under the above assumptions the memory integral
\eqref{MemInt} is governed by Boltzmann behavior. 
Note that this solution remains valid for $H\gg \G_t$,
\ie when the background changes much faster than the time scale $\G_t^{-1}$ associated with the 
dynamics of $\phi$, as long as the much weaker constraint \eqref{smallnesses} holds.
Simplifying the notation, Eq.~\eqref{EffBE} can be
written as\footnote{
Terms of higher order in $f$, such as $f^2-\bar{f}^2$, are contained in (\ref{effBEsimple}) via the $f$-dependence of $\G_t$. They always appear, but are of higher order in the coupling between $\phi$ and other fields because we chose the interaction in (\ref{L}) to be linear in $\phi$ for illustrative purposes. 
} 
\begin{equation}
\partial_t f(t) = -\G_t\left(f(t)-\bar f(t)\right),\label{effBEsimple}
\end{equation} with
\begin{equation}
\bar f(t)\equiv(\G_t^>/\G_t^<-1)^{-1}.
\end{equation}  
If the $\chi_i$ are in local
thermal equilibrium, the gain and loss rates $\G^<_t$ and $\G^>_t$ are
related by the detailed balance condition 
\begin{equation}\label{DetailedBalance}
\G_t^</\G_t^>=e^{-\W_t/T(t)},
\end{equation}
where $T(t)$ denotes an effective temperature; then $\bar f(t)$ becomes
the usual Bose-Einstein distribution.  Note that \eqref{EffBE} and
\eqref{eqn:Dp2} are valid for each field mode $\textbf{q}$ separately.

\section{Discussion}%
The effective kinetic description in terms of the BE (\ref{EffBE}) is well-controlled by the small parameters (\ref{smallnesses}), \ie if all involved time scales are longer than $\TI$. It describes the behavior on the thermodynamic time scales $\Gamma_t^{-1}$, $H^{-1}$, but may receive corrections at times $\lesssim \TI$.  These can be calculated by means of linear response theory. For very large time arguments, the exponential decay behavior of our solution for $\Delta^-$ usually requires corrections \cite{Gautier:2012vh}. 
As both, the spectral properties and our solution for $\Delta^+$, are dominated by contributions from the more recent past, this has little effect on the validity of the effective BE.

The small parameters that fix the accuracy of our WKB solutions for $\Delta^\pm$ are the same as those that control the convergence of the gradient expansion in the Wigner space approach. We therefore expect that the range of applicability of both techniques is similar. 
The advantages of both approaches are, however, complementary.
On one hand, the Wigner space method more closely resembles the diagrammatic expansion of the S-matrix in vacuum, which is also performed in momentum space.
This considerably eases the comparison with results obtained from S-matrix calculations.
This also makes it more suitable to derive an effective Hamiltonian, as e.g. used in neutrino physics \cite{Sigl:1992fn,Canetti:2012vf}. 
It does, on the other hand, require a resummation of infinitely many orders when finite width effects are important in the collision terms \cite{Garbrecht:2011xw}.
Our method includes these effects in an intuitive way without resummation.
It furthermore avoids the Fourier transformation of $\Delta^\pm$, which has to be taken with care in an initial value problem because it requires knowledge of the infinite past and future. 

It is instructive to consider the solutions \eqref{eqn:Ansatz} and
\eqref{eqn:Dp2} locally in time, as they appear \fex in loop diagrams when
calculating self-energies or when deriving properties like the energy density
of fields.  Defining mean and relative times as $t\equiv (t_1+t_2)/2$ and
$y\equiv t_1-t_2$, respectively, we obtain in the limit $y\ll H^{-1}$ that
\begin{align}
  \label{eqn:damping}
  \Dm(t_1,t_2) &\simeq \frac{\sin(y
  \W_t)}{\W_t}e^{-\frac12|y|\G_t}\quad\text{and}\\
  \Dp(t_1,t_2) &\simeq \left[(1+2\bar f(t))e^{-\frac12|y|\G_t} +2\delta f(t) 
  \right]
  \frac{\cos(y \W_t)}{2\W_t}\;,
  \nonumber
\end{align}
where $\delta f(t)\equiv f(t) - \bar f(t)$. The damping rate of the spectral
function as function of $y$ is just given by $\G_t$. The damping of the
statistical propagator can be understood by splitting it up in two parts as
indicated. If the background is in thermal equilibrium, (\ref{DetailedBalance}) is fulfilled and they correspond to the
equilibrium and nonequilibrium parts of the propagator, respectively. In
general, the term proportional to $\delta f(t)$ remains undamped
(\textit{cf.}~discussion in~\cite{Garbrecht:2011xw}). 
The overall damping of $\Dp$ approaches $\G_t$ if $|\delta f(t)|\ll1$. In this case, and
locally in time, our results reproduce the common Kadanoff-Baym ansatz (see \fex Refs.~\cite{Lipavsky:1986zz, Garny:2009ni,
Garbrecht:2011xw, Chou:1984es, Calzetta:1999xh, Greiner:1998vd}).  This
implies that $f(t)$ as defined in \eqref{eqn:DpStep} indeed plays the role of
a generalized phase space distribution function of effective plasma
excitations. Note that $\Gamma_t^\gtrless$ may include off-shell
transport~\cite{Arnold:2000dr, Anisimov:2008dz, Anisimov:2010dk,
Anisimov:2010gy}.

Using \eqref{MemInt}, we can discuss the effect of boundary conditions at
finite time ($t_i=0$ for definiteness). If $\Pp$ is nonsingular at
$t_1=t_2=0$, \eqref{MemInt} implies the initial conditions $\Delta^+_{\rm i}= \dot
\Delta^+_{\rm i}= \ddot\Delta^+_i=0$; here, we defined
$\Delta^+_{\rm i}\equiv\Delta^+|_{t_1=t_2=0}$, $\dot\Delta^+_i \equiv
\partial_{t_1}\Delta^+|_{t_1=t_2=0} = \partial_{t_2}\Delta^+|_{t_1=t_2=0}$ and
$\ddot\Delta^+_i\equiv\partial_{t_1}\partial_{t_2}\Delta^+|_{t_1=t_2=0}$.
Arbitrary initial conditions for $\Dp$ can be implemented by formally adding
$\delta \Pi^+(t_1, t_2) = -\partial_{t_1}\partial_{t_2}\Delta^+(t_1, t_2)
\delta(t_1)\delta(t_2)$ to $\Pi^+(t_1,t_2)$ in \eqref{MemInt}. Boltzmann
behavior according to \eqref{eqn:Dp2} only arises when $\dot\Delta^+_{\rm i}=0$ and
$\W_t\Delta^+_{\rm i}= \W_t^{-1}\ddot \Delta^+_{\rm i} = \frac12 +f|_{t=0}$; otherwise,
\eqref{MemInt} generates oscillating terms that are exponentially damped away
with the rate $\G_t$.

The connection to the classical Boltzmann equation can be made more explicit 
by considering the contribution of the $\phi$-mode $\textbf{q}$ (index suppressed as before) to the energy density, calculated from the energy-momentum tensor\footnote{Here $\upeta={\rm diag}(1,-1,-1,-1)$.} $T_{\mu\nu}^{\phi}(x)=\partial_{\mu}\phi(x)\partial_{\nu}\phi(x)-\upeta_{\mu\nu}\mathcal{L}(x)$, 
\begin{align}\label{Energie}
  \epsilon^{\phi}(t) &= \frac{1}{2}\left(\partial_{t_1}\partial_{t_2} +
  \omega_{t_1}^2
  \right)\left(\Delta^+(t_1,t_2)+\langle\phi(t_{1})\rangle\langle\phi(t_{2})\rangle\right)\big|_{t_1=t_2=t}
\end{align}
Here $\langle\phi(t)\rangle$ is the one-point function or ``mean field'', which in the following we set to zero, and $\omega_t=\sqrt{\q^2 + m(t)^2}$.
For illustration we consider the simplest case, when the background is in thermal equilibrium at constant temperature $T$ and $m(t)$ is a constant.
Then $\G^\gtrless_t$, $\Omega_t$ and $\omega_t$ are constant and (\ref{DetailedBalance}) applies, \ie $\bar{f}=(e^{\Omega/T}-1)^{-1}$. Then (\ref{eqn:Ansatz}) only depends on the relative time, $\Delta^-(t_1,t_2)=\Delta^-(t_1-t_2)$ \cite{Anisimov:2008dz}.
To obtain nonequilibrium behavior, initial conditions $\Delta^{+}_{\text{i}}$,  $\dot{\Delta}^{+}_{\text{i}}$, $\ddot{\Delta}^{+}_{\text{i}}$ for $\Delta^+$ and its derivatives  have to be set up at finite time $t_1=t_2=t_i$ as described above. Then the statistical propagator reads \cite{Anisimov:2008dz} 
\begin{eqnarray}\label{solution}
  \Delta^{+}(t_{1},t_{2}) &=& 
  \Delta^{+}_{\text{i}}
  \dot{\Delta}^{-}(t_{1})\dot{\Delta}^{-}(t_{2})
  +\ddot{\Delta}^{+}_{\text{i}}
  \Delta^{-}(t_{1})\Delta^{-}(t_{2})\nonumber\\
  &+&\dot{\Delta}^{+}_{\text{i}}
  \left(\dot{\Delta}^{-}(t_{1})\Delta^{-}(t_{2})
  +\Delta^{-}(t_{1})\dot{\Delta}^{-}(t_{2})\right)\nonumber\\
  &+& \int_{t_i}^{t_{1}}dt'\int_{t_i}^{t_{2}}dt''
  \Delta^{-}(t_{1}-t')\Pi^{+}(t'-t'')\Delta^{-}(t''-t_{2}),
\end{eqnarray}
where the $\dot{ }$ denotes a time derivative. The last line arises from (\ref{MemInt}) in an obvious way; the other lines are due to the initial conditions at finite time, formally fixed by adding singular pieces to $\Pi^+$. 
Inserting (\ref{solution}) into (\ref{Energie}) with the previous assumptions yields \cite{Drewes:2010pf},
\begin{eqnarray}\label{energiedichtephi}
  \lefteqn{\epsilon^{\phi}(t)=\frac{\Delta^{+}_{\text{i}}}{2}
  \left(\frac{\omega^{2}-\Omega^{2}}{2}\cos(2\Omega\tp)+\frac{\omega^{2}+\Omega^{2}}{2}\right)e^{-\Gamma\tp}}\nonumber\\
  &-&\frac{\ddot{\Delta}^{+}_{\text{i}}}{2\Omega^{2}}
  \left(\frac{\omega^{2}-\Omega^{2}}{2}\cos(2\Omega\tp)-\frac{\omega^{2}+\Omega^{2}}{2}\right)e^{-\Gamma\tp}\nonumber\\
  &+&\frac{\dot{\Delta}^{+}_{\text{i}}}{\Omega}
  \frac{\omega^{2}-\Omega^{2}}{2}\sin(2\Omega\tp)e^{-\Gamma\tp}%\nonumber\\
   + \left(\frac{1}{2}+\bar{f}\right)\frac{\omega^{2}+\Omega^{2}}{2\Omega}\left(1-e^{-\Gamma\tp}\right).
\end{eqnarray}
The BE describing the classical analogue of this system - a collection of $\phi$ particles coupled to a large thermal bath - would read $\partial_{t}f(t) = -\Gamma(f(t)-\bar{f})$, with the solution $f(t)=\bar{f}+(f_{\text{i}}-\bar{f})e^{-\Gamma t}$ and $\epsilon^{\phi}(t)=\omega f(t)$. In these expressions $f(t)$ is the classical particle number and $f_{\text{i}}$ its initial value.
The general shape of (\ref{energiedichtephi}), including oscillations in time, is very different. 
This is expected because the quantum system can be prepared in a state that does not correspond to a definite particle number. 
If the initial conditions are chosen as 
%\begin{center}
 % \begin{tabular}{c c c}
    $\Delta^{+}_{\text{i}}  
    = \frac{1}{\Omega}
    \left(\frac{1}{2}+f_{{\rm i}}\right)$, %& 
$\dot{\Delta}^{+}_{\text{i}}  
    = 0 $, %& 
$\ddot{\Delta}^{+}_{\rm i}  = \Omega
    \left(\frac{1}{2}+f_{{\rm i}}\right)$,
%  \end{tabular}
%\end{center}
one obtains an expression that (up to a vacuum energy) reproduces the classical expression in the limit $\Omega\rightarrow \omega$,
\begin{eqnarray}
  \epsilon^{\phi}(t)=\frac{\Omega^{2}+\omega^{2}}{2\Omega}\left(
  \left(\frac12+\bar{f}\right)+\big(f_{{\rm
  i}}
  -\bar{f}\big)e^{-\Gamma t}\right).\label{tsunami}
\end{eqnarray}
Note that the vacuum piece depends on the background properties (e.g. temperature) for $\Omega\neq\omega$. Similar terms containing combinations $\Omega^2\pm\omega^2$ also appear in the expression for the pressure and can lead to a negative equation of state for $\Omega^2<\omega^2$ \cite{Anisimov:2008dz}.

In the early universe, quantum fields propagate in an expanding background
described by a Friedmann-Robertson-Walker metric with scale factor $a(t)$ and
Hubble rate $H\equiv \dot a/a$. Since in conformal coordinates this is
equivalent to a time-dependent mass term and self-energies, expansion is already accounted for
in our calculation. The BE \eqref{EffBE} remains unchanged, but the
distribution function becomes a function of comoving momentum
$\q_\text{com}=a(t)\q$. Going back to units of physical space and momentum
yields the well-known result
\begin{align}
  \label{EffBE2}
  \partial_t f_\q(t) = \left( 1+f_\q(t) \right)\G^<_{t}\! -\! f_\q(t)\G^>_{t}
  +H\q \nabla_\q f_\q(t) \;.
\end{align}

\section{Conclusions}%
We have presented a simple method to derive effective Boltzmann equations (BEs) from the fundamental 
Kadanoff-Baym equations (KBEs), and
illustrated it in case of a single real scalar field $\phi$. 
In contrast to previous derivations, which are based on a gradient expansion of the KBEs in Wigner space, we derived approximate WKB
solutions to the full KBEs.
These are valid under the very general physical assumptions of
weak coupling and separation of macroscopic and microscopic time scales.
This includes situations where the background evolves much faster than $\phi$.
The statistical propagator can be expressed in terms of a generalized distribution function that follows a generalized BE, \ie a first order differential equation that is local in time. 
The accuracy of the BE is controlled by the accuracy of the WKB solution to the full KBE.
Locally in time, our solutions for the correlation functions reproduce the common Kadanoff-Baym ansatz when the deviation from equilibrium is small.
The presented approach can be extended to fermions with gauge interactions and 
multi-flavor problems, including flavor oscillations, 
as well as beyond the weak damping regime. 
Boltzmann behavior arises whenever the WKB approximation is justified. \\ \\

\textbf{\large Acknowledgments} - We are very grateful to J\"urgen Berges, Mathias Garny, Georg Raffelt and Julien Serreau for useful discussions and comments on the manuscript.
This work is supported by the Gottfried Wilhelm Leibniz programme of the Deutsche Forschungsgemeinschaft. C.W. acknowledges partial support from the European 1231 Union FP7 ITN INVISIBLES (Marie Curie Actions, PITN-GA-2011-289442).

\bibliographystyle{JHEP}

\end{document}